# Non-invasive urinary bladder volume estimation with artefact-suppressed bio-impedance measurements

Kanika Dheman*[1,2], Stefan Walser[2], Philipp Mayer[3], Manuel Eggimann[3], Marko Kozomara[4], Denise Franke[4], Thomas Hermanns[4], Hugo Sax[5], Simone Schürle*[6] and Michele Magno*[1]

*Abstract*— Urine output is a vital parameter to gauge kidney health. Current monitoring methods include manually written records, invasive urinary catheterization or ultrasound measurements performed by highly skilled personnel. Catheterization bears high risks of infection while intermittent ultrasound measures and manual recording are time consuming and might miss early signs of kidney malfunction. Bioimpedance (BI) measurements may serve as a non-invasive alternative for measuring urine volume in vivo. However, limited robustness have prevented its clinical translation. Here, a deep learning-based algorithm is presented that processes the local BI of the lower abdomen and suppresses artefacts to measure the bladder volume quantitatively, non-invasively and without the continuous need for additional personnel. A tetrapolar BI wearable system called ANUVIS was used to collect continuous bladder volume data from three healthy subjects to demonstrate feasibility of operation, while clinical gold standards of urodynamic (n=6) and uroflowmetry tests (n=8) provided the ground truth. Optimized location for electrode placement and a model for the change in BI with changing bladder volume is deduced. The average error for full bladder volume estimation and for residual volume estimation was -29±87.6 ml, thus, comparable to commercial portable ultrasound devices (Bland Altman analysis showed a bias of -5.2 ml with LoA between 119.7 ml to -130.1 ml), while providing the additional benefit of hands-free, non-invasive, and continuous bladder volume estimation. The combination of the wearable BI sensor node and the presented algorithm provides an attractive alternative to current standard of care with potential benefits in providing insights into kidney function.

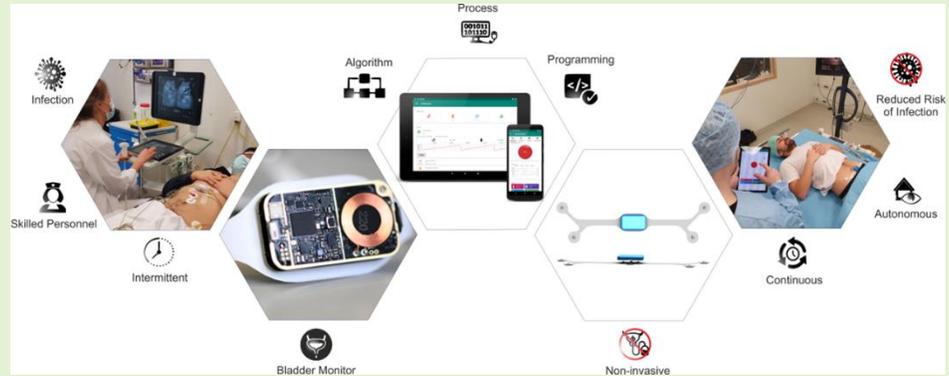

*Index Terms*—non-invasive, bladder monitoring, wearables, machine learning, bio-impedance

## I. Introduction

Urine volume production is a critical marker of kidney function and a vital parameter that can be used to monitor health [1]. A large portion of the patient population requires the measurement of urine volume in the bladder [2]. This includes patients with spinal cord injury, patients in intensive, intermediate, and post-operative care, patients with trauma in emergency and chronic diseases such as diabetes and acute kidney injury. Reduced urine output has been shown to detect acute kidney injury earlier than increased serum creatinine level [3], [4]. In clinical settings, urinary catheterization is the most prevalent method of measuring urine volume. Clinical staff manually record the urine volume every 3 to 4 hours from catheter-attached urine bags, proving only intermittent and low granularity changes in urine output. A potent risk factor linked with catheterization of the urinary tract is catheter associated urinary tract infection (CAUTI), which is the most common healthcare associated infection that has been shown to lead to an increase in direct and indirect hospital costs, morbidity, and mortality [5], [6]. Also, urinary catheterization is a major cause of non-infectious urogenital complications such as mechanical trauma to the lower urinary tract[7]. Another complication is bacteriuria which often leads to the unnecessary use of antimicrobial agents, making urinary drainage systems reservoirs for multidrug-resistant bacteria and a source of transmission to other patients [8]. The USA Centers for Disease Control and Prevention (CDC) guidelines strongly advocate minimizing the use and the duration of inserted catheters to prevent CAUTI [9], [10] and stipulate that catheterization of the urinary tract is carried out only when necessary. On the other hand, ultrasound measures of the bladder to measure urine volume non-invasively are intermittent, in need of infrastructure and subjective to the skill and interpretation of a trained clinical professional. All methods implemented in current clinical practice have the following drawbacks: temporally sparse data that might miss early signs of kidney malfunction, risk of infection during urinary catheterization and increased workload for the clinical staff with demands on both resources and time [7]. Hence, there is a need for a non-invasive method for measuring the urine volume in the bladder that reduces the risk of infection and also detects early signs of kidney malfunction by providing accurate, autonomous, and continuous bladder volume measurements [11], [12]

[1]Project Based Learning Center, ETH Zürich, Switzerland [2]Multi-Scale Robotics Lab, ETH Zürich, Switzerland. [3] Integrated Systems Laboratory, ETH Zürich, Switzerland. [4]Department of Infectious Diseases, Bern University Hospital, University of Berm, Switzerland. [5]Klinik für Urologie, Unispital Zurich, Switzerland. [6] Responsive Biomedical Systems Laboratory, ETH Zurich, Switzerland

.The change in the BI of the lower abdomen could be a promising approach to circumvent the aforementioned challenges and provide a method for non-invasive, automated (hands-free) and continuous urine volume measurement [13]–[15]. BI is a property of all tissue substrates that can conduct electrical current which has been extensively researched [16], [17], and used for non-invasive, safe and low cost assessment of body composition for both at-home and clinical applications. The method has especially been studied for scenarios where the body's fluid composition changes over time, such as, dehydration and edema [18]. However, BI measurements are still not optimized for continuous use and suffer from artefacts due to external disturbances such as motion, pressure and temperature [19]. Under such conditions it is difficult to obtain accurate measurement of urine volume in-vivo.

In this work, an algorithm for a wearable BI system, **ANUVIS** (**A**utomated **N**on-invasive **U**rine **V**olume for **I**n-vivo **S**ensing) is presented that can measure urine volume quantitatively at a high temporal resolution, non-invasively and without the need for additional personnel. The ANUVIS sensor node measures the local bio-impedance of the lower abdomen with only four electrodes and is reported previously [19]. In this work, the ANUVIS system is further developed for non-invasive bladder monitoring where the measured BI data is processed with a low complexity algorithm to provide an artefact-corrected, accurate, and quantifiable bladder volume estimate. The combination o f a low-power wearable sensor node along with an accurate, low-complexity algorithm with artefact-corrected BI measures could surpass the limitations of current clinical procedures (invasive catheterization and intermittent manual record keeping and US measurements by a skilled personnel) by providing automated, non-invasive measures with high temporal resolution and enhanced patient comfort.

The specific contributions of this work are:
• Evaluation of the change in the BI of the lower abdomen during bladder filling and voiding to optimize the placement of the electrodes.
• Deduction of a model for change in BI of the lower abdomen during change in the bladder volume (measured with urological gold standards for bladder volume assessment: urodynamic tests and uroflowmetry tests).
• Development of an algorithm to process the BI measurements in order to provide quantifiable bladder volume estimation with artefact identification and correction.
• Demonstration and evaluation of the sensor system for bladder volume estimation with repeated measures on healthy adults.

## II. RELATED WORK

Different methods have been evaluated for non-invasive measurement of the urine volume in the bladder such portable and wearable ultrasound, electrical impedance tomography, and magnetic resonance imaging[11]. Ultrasonic (US) bladder volume sensing was first used and tested on healthy human population as a portable, pocket-size sensing system [20]. This sensor gave a qualitative binary decision on the bladder volume of either full or empty. Recent research has tried to *quantify* the volume in the bladder with portable/handheld US measurements. A clinical study assessed the performance of two commercially available portable US bladder scanners namely, the *BladderScan® BVI 9400* [21] and *the Prime®* [22] against catheter acquired urine volumes for measuring the post operative urine retention in 318 surgical patients[23]. The BVI 9400 overestimated the bladder volume by 17.5% while the *PrimePlus* system underestimated the volume by 6.3%. Similarly, other studies have highlighted the inaccuracy of bladder volume estimation with portable/handheld bladder scanners to bedside ultrasound machine in intensive care units[24], [25]. In all cases of using US as a method for bladder volume estimation, measures were taken intermittently every few hours with the portable/handheld device by the clinical staff. None o f the US systems were able to provide automated (hands-free) and quantitative bladder volume measurement without the assistance of skilled professional.

To surpass these limitations of US systems the feasibility of using non-invasive bio-impedance measurements was evaluated.[14], [15] Multi-electrode systems for electrical impedance tomography (EIT) measure the electrical impedance around the waist to form an image of the intravesical changes as urine fills and voids the bladder. A 16 electrode EIT belt was used to measure the change in impedance in real time with fluid inflow during urodynamic testing and showed a negative co-relation of the impedance value with the bladder volume [15][26]. However, some subjects showed a positive co-relation that was not investigated further, nor was a method to quantify the bladder volume was provided. Furthermore, the authors suggested that future investigations should study different electrode positions. Different electrode positions during an EIT measurement with 16 electrodes was studied in [27] and it was reported that a grid placement of electrodes on the ventral surface of the lower abdomen performed better than a belt-like placement of electrodes across the pelvic region. Additionally, major measurement problems due to movement artefacts and abdominal muscle contraction were shown to cause volume independent changes in the impedance measured, making real time monitoring difficult. Another study also used an EIT belt with 16 electrodes around the waist on five immobile human subjects and reported a mean volume estimation error of 22.5% (47.2% for 100 ml and 10.6% for 500ml) [26]. Supervised machine learning classifiers have also been used to determine a binary state of bladder, full or not full, from electrical impedance measurements [28]. Deep neural networks (DNNs) have been trained to estimate urinary bladder boundary with EIT belts on a synthetic bladder phantom and showed better performance than traditional approaches in estimating the bladder volume [29]. Despite all these promising efforts, EIT depends on a multi-electrode configuration (n ≥ 8) that measures the spatial variations in potential inside a biological substrate and are complex for wearable application. Due to this challenge, these strategies might face major limitations in terms of artefacts and usability in both clinical and home environments.

Tetrapolar (n=4) BI measurement systems greatly reduce system complexity. Prior work has investigated the feasibility of tetrapolar single-frequency BI measurements to be able to estimate full bladder [14], bladder volume in animals [30], children [31], and healthy adults[32]. However, a method to obtain quantifiable volume of urine in the bladder of human subjects converting the measured changes in BI has not been reported. In addition, the exact electrode locations, the separation between the electrodes or the evaluation of repeated measures within the same subject, has not been described in previous investigations.

The aim of this work is to account for the artefacts in BI measurements to develop an application for non-invasive bladder volume estimation. This aim is achieved by developing an algorithm that interprets the tetrapolar BI measurements of the lower abdomen (measured with a single frequency (50kHz) low-power wearable sensor node [19]), identifying and correcting artefacts to provide hands-free quantitative measures of the bladder volume and without the continuous presence of skilled professionals.

## III. EXPERIMENTAL METHODS

Data was collected during bladder filling and bladder voiding. The ground truth on bladder volume was provided by urological gold standards for bladder volume assessment: urodynamic tests (bladder filling) and uroflowmetry tests (bladder voiding). In total, 14 tests (8 cases of bladder voiding and the 6 cases of bladder filling) are conducted on healthy adults (two Male and one Female) with at least 2 repeated measures on each subject. Regulatory approval from a local ethics body for study number ETH-EK, 2019-N-145. Informed consent, both written and verbal, was taken from all participants.

### A. Setup and Materials

A tetrapolar BI sensor node: ANUVIS, that was developed in a previous work was used for data collection [19]. The system was configured to have a source excitation frequency of 50kHz sampled every 300ms for temporally dense data collection. The Ambu Bluesensor R wet electrodes[33] were used as transducers. The ground truth on the bladder volume during filling was provided by the *NXT Pro Advanced Urodynamic System by Laborie*.[34] While during the bladder voiding experiments, a uroflowmeter *MMS Flowmaster by Laborie* [35] was used.

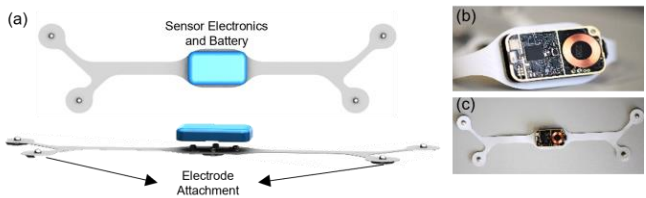

Fig 1. (a) Design of the ANUVIS wearable showing the placement of the sensor electronics, battery and sites for electrode attachment [19]. (b) ANUVIS sensor electronics and (c) assembled wearable for bladder volume measurement

### B. Testing Protocol

1. During the urodynamic and uroflowmetry testing, the ANUVIS system was worn by the subject to acquire BI data in parallel. Four Ambu R electrodes were placed on the ventral surface of the lower abdomen to measure the BI. Prior to conducting the urodynamic and uroflowmetry tests, the intravesical volume of the bladder was studied with US for full bladder volume and residual bladder volume to ascertain the placement of the four electrodes. This was done to ensure that the full range of bladder volume change could be evaluated.

2. Each urodynamic test was carried out in a clinical setting with two consecutive bladder filling cycles, where saline is injected into the urinary bladder with a catheter at a defined flowrate.

3. After the bladder filling process, the subject was asked to void the saline solution. The decreasing volume of the

bladder was measured with the uroflowmeter.

## IV. ALGORITHM

BI data is susceptible to artefacts due to externally caused disturbances (motion, applied pressure and temperature). It is necessary to be able to correctly interpret the cause for the change in measured BI as artefact or intravesical volume change to enable continuous monitoring of the bladder volume Hence, the presented algorithm has four main stages of execution: a) Stage 1: data processing and feature extraction, b) Stage 2: artefact identification, c) Stage 3: artefact suppression and d) Stage 4: conversion of change in BI to bladder volume in milliliters. If no artefact is detected, the data is directly passed to Stage 4 for conversion of bio-impedance to volume. The algorithm was developed and implemented on the collected BI data offline using MATLAB 2020b software.

### A. Data Processing and Feature Extraction (Stage 1)

The incoming time series data is sectioned into windows of predefined length. Bladder filling (BF), being a slow process (taking several minutes during urodynamic testing and many hours during natural human function) is analyzed in windows of 30s. While bladder voiding (BV), which is a faster process that can last up to only 1-2 minutes is analyzed in 1 s windows. The temporal data in each window is smoothed using the robust locally weighted scatterplot smoothing filter with a first-degree polynomial model that assigns a lower weight to the outliers in the regression. Data points in each window are normalized to give a z-score that accounts for inter-subject variability. Also, the baseline noise differs across subjects due to variance in the skin electrode contact (due to skin physiology, body hair, keratinization, moisture and ambient conditions)[36]. In order to correctly identify artefacts, the first data window is used to calibrate the system with respect to the baseline noise.

Descriptive statistical features are extracted for each data window. Eleven BI features and eleven SE impedance features for each of the four electrode contacts were extracted. These included the mean, standard deviation, minima, maxima, energy, entropy, gradient of the mean of successive windows, drift flag, slope, and intercept of linear fit. Before commencing *Stage 2* of artefact identification, the features were ranked to identify the most relevant features for identifying and classifying artefacts. The recursive feature extraction (RFE) and the maximum relevancy minimum redundancy (MRMR) algorithms were used. The results of these are reported in Table A and Table B respectively of the supplementary material. Since the outputs of both algorithms differed, features with high relevance scores were grouped into feature sets as described in Table I. Each of these feature sets were used as input vectors for machine learning classifiers (in *Stage 2*) to evaluate the optimum input vector for the task of artefact identification and classification.

TABLE I
FEATURE COMBINATIONS THAT ARE EVALUATED FOR CLASSIFICATION OF ARTEFACTS IN BI DATA

| ID | #Features | Description of feature set |
|----|-----------|---------------------------|
| 1 | 8 | Highest ranked BI features with RFE algorithm |
| 2 | 11 | Highest ranked BI features with MRMR algorithm |
| 3 | 14 | BI and all SE features ranked highest by RFE |
| 4 | 12 | BI and all SE mean features ranked highest by RFE |
| 5 | 10 | BI and all SE energy features ranked highest by RFE |

### B. Artefact Identification ( Stage 2)

<u>Artefacts Labels:</u> Previous work has identified specific characteristics of artefacts in BI data due to motion, applied pressure, changes in skin temperature and posture variations in the upper body BI [13], [37]. These were: high variance noise where the variance of the data window is 2.5 times greater than that at calibration, drifts between data windows and a combination of the previous two characteristics[13].Thus, four labels were identified as : no change in BI (L=0), high variance noise ( L=1), positive drift( L=2) and negative drift (L=3). The dataset [37] was used to train machine learning classifiers of support vector machines(SVM) and deep neural networks (DNN) offline to identify the characteristics in the data window.

<u>Machine Learning Classifiers:</u> The SVM was trained and optimized by performing a grid search to provide the best fit parameters: radial basis function as the kernel, Gamma = 0.1, C = 100. Similar to SVM classifier, a grid search was performed for parameter optimization for the DNN. The best results were obtained with a network of a fully connected structure with 6 hidden layers and 8 nodes each. The activation function *relu* was used for all layers of the DNN except for the output layer where the sigmoid function was applied. Data was split into k=10 equally sized parts or folds to perform a k-fold cross validation to evaluate the classifier performance.

Figure 2 shows the weighted accuracy of all labels (L=0,1,2,3) for different feature combinations (as described in Table I) used as input vectors. Out of all the combination groups in Table I, feature set (ID-4) performed the best followed by feature set (ID-3) for both the SMV and DNN classifier. However, the feature set (ID-1) is only 0.9% lower in accuracy than the top performer and has lower dimensionality than features sets (ID-3) and (ID-4). both these feature sets are resource intensive, which could be a limiting factor for future online implementation. Hence, for further training and evaluation the feature set (ID-1) was used because it was the smallest in size while maintained classification performance similar to larger feature sets, implying less computational load while maintaining performance.

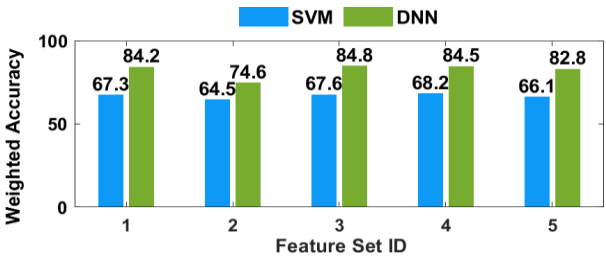

Fig 2. Comparison of the accuracy score reported by the hyper-parameter tuning for the SVM and DNN with the feature combinations chosen in TableI

### C. Artefact Correction ( Stage 3)

*Context Awareness:* In order to correctly suppress an identified artefact, it is necessary to understand the context of measurement. There are two main events to be monitored by the presented algorithm: bladder filling and bladder voiding. Bladder filling is a relatively slower physiological process when compared to bladder voiding. Hence the algorithm considers the default state of measurement to be in bladder filling, that is, the

BI data should be decreasing with time. An aberration to this trend would be considered an artefact and suppressed as discussed below. Bladder voiding is detected with inertial sensors and an increase in bio-impedance over a few minutes. In this context, a decrease in BI is considered an artefact that needs to be suppressed.

Artefacts in BI data during bladder filling and voiding processes are suppressed using a 2-D Kalman filter. Experiments showed that the change in BI with changing bladder volume could be represented as a linear model with negative slope, i.e., since the BI decreases with increasing bladder volume. This model was used for artefact suppression while implementing the 2-D Kalman filter. Additionally, healthy adults produce urine at a rate of at least 1ml per kg of body weight hour. and is dependent on the glomuler filtration rate. Thus, in the analyzed data window of 30s, a constant rate of change of bladder volume was assumed.

*Variable Initialization:* The first data window is used for system calibration and initialization of variables representing the estimates used in the 2-D Kalman filter for artefact suppression The initial bladder volume is assumed to be 0 ml. The initial BI

---

**Algorithm 1: Bladder volume estimation from BI data**

| **Input** | : for *n* data points per parameter in matrix $X^i$ |
|---|---|
| **Output** | : Bladder volume estimation in milliliters |
| **Feature Extraction** | : Data normalisation,standardisation and feature extraction |

*if* (artefact identified == TRUE)
*Volume Estimation for window 'i'*
**Update Equations** :

$$G_{BI}(i) = \frac{\Gamma_{\widehat{BI}}(i)}{\Gamma_{\widehat{BI}}(i) + \Gamma_{BImeasure}(i)} \quad \text{..............................(1)}$$

$$G_{BI\prime}(i) = \frac{\Gamma_{\widehat{BI}\prime}(i)}{\Gamma_{\widehat{BI}\prime}(i) + \Gamma_{BI\prime measure}(i)} \quad \text{..................(2)}$$

$$\widehat{BI}(i) = \widehat{BI}_p(i) + G_{BI}(i).(BI(i) - \widehat{BI}_p(i)) \quad \text{.................(3)}$$

$$\widehat{BI}\prime(i) = \widehat{BI}\prime_p(i) + G_{BI\prime}(i).(BI\prime(i) - \widehat{BI}\prime_p(i)) \quad \text{.........(4)}$$

$$\Gamma_{\widehat{BI}}(i+1) = \Gamma_{\widehat{BI}}(i).(1 - G_{BI}(i)) \quad \text{...........................(5)}$$

$$\Gamma_{\widehat{BI}\prime}(i+1) = \Gamma_{\widehat{BI}\prime}(i).(1 - G_{BI\prime}(i)) \quad \text{..........................(6)}$$

$$\Delta\widehat{BI} = \widehat{BI}(i) - \widehat{BI}(i-1) \quad \text{...........................(7)}$$

$$\Delta V(i) = \delta.\Delta BI(i) \quad \text{...........................(8)}$$
*Volume (ml) aggregated till window 'i'*

$$V(i) = cumsum(\Delta V(1:i)) \quad \text{...........................(9)}$$
**Prediction Equations** :Predict estimate for the next state in window 'i+1'

$$\widehat{BI}_p(i+1) = \widehat{BI}(i) + \Delta(t).\widehat{BI}\prime(i) \quad \text{...........................(10)}$$

$$\widehat{BI}\prime_p(i+1) = \widehat{BI}\prime(i) \quad \text{...........................(11)}$$

$$\Gamma_{\widehat{BI}}(i+1) = \Gamma_{\widehat{BI}}(i) + \Delta(t)^2.\Gamma_{\widehat{BI}\prime}(i)) \quad \text{..................(12)}$$

$$\Gamma_{\widehat{BI}\prime}(i+1) = \Gamma_{\widehat{BI}\prime}(i)) \quad \text{...........................(13)}$$
*end if*

---

Fig. 3. Algorithm for artefact corrected single frequency BI data for bladder volume estimation using a wearable tetrapolar BI sensor node.

estimate $(\widehat{BI})$ is set to the mean of the first data window $(\mu_C)$. The rate of change of the bladder volume is assumed to be 1ml/min. and the initial rate of change of the BI estimate between windows is set to the sensitivity parameter (δ). δ is defined as defined as the ratio of the absolute change in lower abdomen BI to the absolute change in bladder volume in units of in Ω/ml. The initial measurement uncertainty for measured BI ($\Gamma_{BIinitial}$) with the variance of the measurements in the calibration data window ($\sigma^2$). Assuming high error in estimation, a high initial estimate uncertainty ($\Gamma_{BIinitial}$) of 100 $\Omega^2$ is assumed. This is used to calculate the first Kalman Gain after initialization.

*Implementation of artefact suppression:* Artefacts identified in the previous *Stage 2* of the algorithm are suppressed as shown in Algorithm 1in Figure 3. For every system state *'i'* the filter applies the weight or gain factor ($G_{BI}$) that is calculated with previous measurement and estimate uncertainties (Figure3: Algorithm 1, lines (i) to (ii)). Hence, in a given data window with an artefact, an increase in the measurement uncertainty results in a lower value of the Kalman Gain. The current system state estimate and the estimate uncertainty can then be updated as given in equations (iii) to (vi). After the estimates for BI and BI' for the current state *'i'* are calculated, the prediction for the next data window *'i+1'* is made according to the model depicting system behavior, represented by equations (x) to (xiii).

### D. Volume Estimation (Stage 4)

The volume of fluid in the bladder is computed with the relative changes between the estimates $\widehat{BI}$ of consecutive data windows, as given by equation (viii) and (ix) of Algorithm 1. The volume increments (during bladder filling) and the volume decrements (during bladder voiding) are calculated using the δ as a conversion factor equating change in BI to change in bladder volume for each analyzed data window. The total bladder volume or voided volume is calculated as a cumulative sum of all windows analyzed up to the current window *'i'*.

## V. RESULTS

The test for measuring bladder volume in-vivo were conducted on three subjects with demographics provided in Table II. A total of six urodynamic tests were performed. Two consecutive bladder filling cycles were conducted with on subject 1 (S01: n= 2) and four urodynamic tests were conducted with subject 2 (S02: n=4). While eight uroflowmetry tests were done on all subjects (S01: n=2, S02: n=4 and S03: n=2).

### A. Electrode Placement

Fig. 4. Electrode placement for tetrapolar BI measurement on the lower abdomen.

An exploratory investigation of the intravesical changes of the bladder with ultrasound scans showed that the bladder expanded in the coronal plane laterally towards the pelvic bones, towards the navel from the pubic bone in the sagittal plane and in depth towards the spine. To measure the full dynamic range of bladder expansion and contraction, the electrodes were placed such that the vertical, horizontal and in-depth volume changes in the bladder could be recorded. Hence, the electrodes were placed as ss shown in Figure 4. The source electrode was placed over the pubic bone, 4cm below the navel. The positive voltage measurement electrode was placed 3cm to the right of the source electrode along the same lateral axis. The sink electrode (system ground) was placed adjacent to the navel on the left side. The negative voltage measurement electrode was placed 3cm to the left of the sink electrode (system ground) along the lateral axis. Such an electrode placement enabled tracking the changes in the bladder volume from the full bladder to the empty bladder (with residual urine volume during the natural voiding process.). In addition, repeated measurements during urodynamic and uroflowmetry tests and a linear model for BI change with bladder volume were also deduced with this electrode placement.

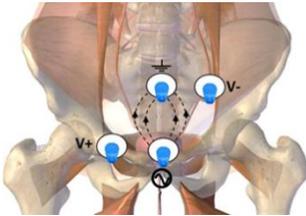

### B. Artefact Identification

The machine learning classifiers can accurately identify different artefacts and their associated characteristics. Figure 5 plots the receiver operating characteristic (ROC) curve of the SVM classifier for all 4 Labels. The area under the ROC curve (AUC) is a direct measure of the diagnostic ability of the classifier. Label 0 (no change in BI) and Label 1 (high variance noise) have high AUC values of over 0.96 while the AUC values of Labels 2 (positive drift) and 3 (negative drift) are lower at 0.87.

Figure 6 shows the ROC-curve for all the labels of the DNN.

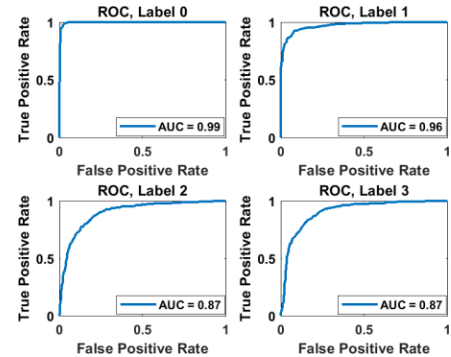

Fig. 5. ROC plots for all labels of SVM classifier

Label 0 has a high AUC value of 0.99 with very high sensitivity values at low false positive rates. With an AUC value of 0.91, Label 1 classification also shows high sensitivity at lower specificity values. As before, the two drift Labels 2 and 3 have similar characteristics as for SVM with an AUC value of 0.86.

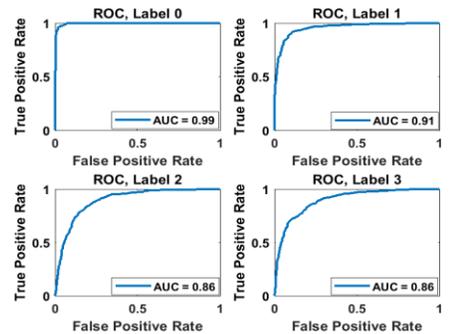

Fig. 6. ROC plots for all labels of DNN

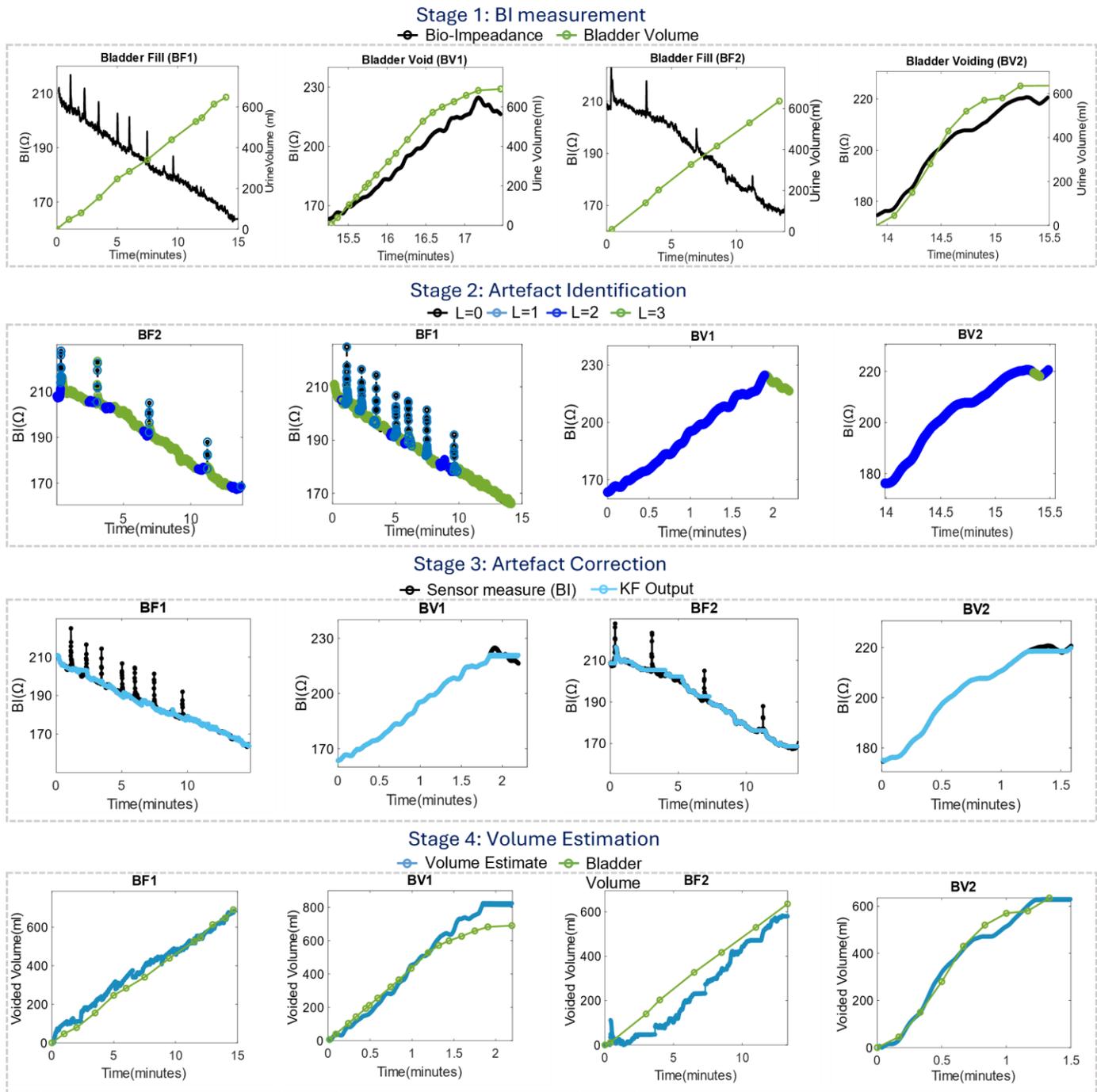

Fig. 7. Four stages of bladder volume estimate algorithm showing both bladder filling cycles (BF1 and BF2) and bladder voiding cycles (BV1 and BV2) during urodynamic and uroflowmetric tests respectively, on subject S01. (a) Stage 1 of the algorithm with raw BI data measured as the bladder expands and contracts, followed by (b) Stage 2 of artefact identification where four features of no change (L=0), high variance noise (L=1), positive drift (L=2) and negative drift (L=3) are identified , (c) Stage 3 for artefact correction given the context of measurement and (d) Stage 4 for quantitatively estimating the bladder volume .

## C. Bladder Filling With Urodynamic Tests

Six urodynamic tests with repeated measures on subjects were performed on two subjects (S01, S02). Figure 7 (BF1, BF2) shows the change in the BI of the lower abdomen of S01 as the bladder is filled with saline solution during urodynamic testing. The BI decreased monotonically with the increase of saline in the bladder, elucidating the linear model of BI change with bladder volume. For S01, measured impedance at the start of the bladder filling was same for both BF1 and BF2, indicating repeatable measurements when there is no residual urine in the bladder. The filling of the bladder with saline solution during the test was stopped when the subject expressed the feeling of bladder "fullness" to the urologist. The two FILLING cycles recorded an absolute decrease of 47.9Ω for a volume of 691ml (Figure 7 (BF1)) and an absolute decrease of 40.2Ω for a volume change of 636ml (Figure 7 (BF2)). Periodic spikes in bio-impedance were observed at the instances when the subject was asked to cough by the clinician during urodynamic testing, showing an influence of local motion and muscle activity on the measured bio-impedance.

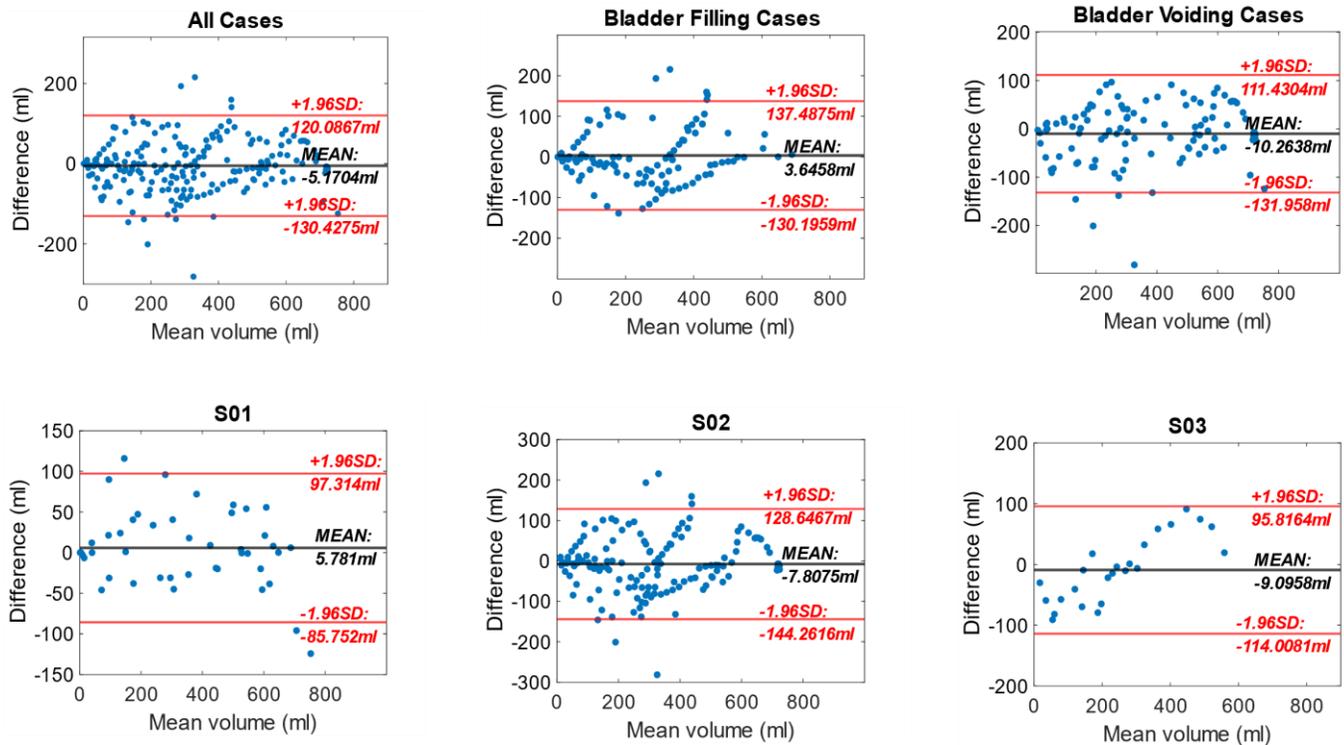

Fig. 8. Bland Altman plots showing the error during the estimation of the bladder volume using the presented algorithm for (a) all 14 instances of bladder filling and voiding, (b) 6 cases of bladder filling, (c) 8 cases of bladder voiding, intra-subject variability for (d) S01, (e) S02 and (f) S03

### D. Bladder Voiding With Uroflowmeter Tests

During bladder voiding, an increase in BI of the lower abdomen with the evacuation of urine from the bladder was observed for all the cases. Figure 7 (BV1 and BV2) show the absolute change in BI of the lower abdomen of subject S01, measured during bladder voiding with a catheter and ground truth provided by the uroflowmeter. In S01, an absolute increase of 52.5Ω is observed for voiding of 691ml of saline (Figure 7 (BV1)). and 47.92Ω for a bladder volume of 636ml (Figure 7 (BV2)). With respect to the filling cycle BF1 and BF2, this is an increase of 4.5Ω and 7.7 Ω respectively, for same change in bladder volume. This increase is seen in the BI at the end of the voiding process, where tensing of the muscle might have resulted in a change in muscle tenor and thereby muscle BI [27].

### E. Bladder Volume estimation

Figure 7 shows the successive stages of artefact detection, correction and volume estimation executed by the bladder volume estimation algorithm on the bladder filling and voiding data for S01. It can be seen that all features specified for the deep learning classifiers were accurately identified in both the bladder filling and bladder voiding cases. Artefacts were identified with respect to context and were suppressed in accordance to the used linear model for BI change with changing bladder volume. Section IV of the supplementary material shows the propose algorithm implemented for all cases.

### F. Error Analysis

Table II summarizes the error ($\Delta$) with respect to the ground truth in the total bladder volume at the end of the bladder filling and voiding cycles for each measurement case. The average error in estimating the total bladder volume, full bladder for cases of bladder filling and residual volume for bladder voiding, was calculated to be -29.65 ± 87.6ml.

However, to evaluate the performance of this sensor system as a means to provide real time unobtrusive bladder volume

TABLE II
ERROR IN TOTAL BLADDER VOLUME AT THE END OF BLADDER FILLING AND VOIDING

| CASE | S-ID* | F/M | Age | GT | EV | $\Delta$ |
|------|-------|-----|-----|------|---------|--------|
| BF1 | S01 | M | 65 | 691 | 685 | -6 |
| BF2 | S01 | M | 65 | 636 | 580.3 | -55.7 |
| BF3 | S02 | F | 31 | 509 | 367.9 | -141.1 |
| BF4 | S02 | F | 31 | 438 | 222.395 | -215.6 |
| BF5 | S02 | F | 31 | 511 | 521.823 | 10.82 |
| BF6 | S02 | F | 31 | 517.5 | 357 | -160.5 |
| BV1 | S01 | M | 65 | 691 | 815.32 | 124.32 |
| BV2 | S01 | M | 65 | 635 | 627 | -8 |
| BV3 | S02 | F | 31 | 717.4 | 728.746 | 11.34 |
| BV4 | S02 | F | 31 | 553.6 | 540.486 | -13.11 |
| BV5 | S02 | F | 31 | 317 | 290.875 | -26.12 |
| BV6 | S02 | F | 31 | 246 | 331.298 | 85.29 |
| BV7 | S03 | M | 32 | 568.4 | 546.272 | -22.12 |
| BV8 | S03 | M | 32 | 307.9 | 309.326 | 1.426 |

* Subject ID; F: Female; M: Male; GT: Ground Truth(ml); EV: Estimated Volume(ml); $\Delta$: Error(ml)

without the need for an expert, an analysis was conducted at multiple intermediate points during the bladder filling and voiding cycles. Figure 8 shows the Bland Altman plots to analyze the error in the sensor node measured at multiple timestamps to compare with the ground truth. Overall, as seen in Figure 8(a), a bias of -5.2ml with limits of agreement (LOA) between 119.7ml to -130.1ml was observed. However, due to the unforeseen effects, probably due to muscle activity and change in muscle tenor, the accuracy of continuous bladder estimation was lowered. This can be seen for both the urodynamic (Figure8(b)) and uroflowmetry (Figure8(c)) tests. Intra-subject variability within S01 shown in Figure8(d), revealed a positive bias of 5.78ml with LOA within 100ml. While S02 (Figure 8(e)) and S03 (Figure 8(f)) both showed negative bias of -7.8ml and -9 ml respectively. Repeated tests on S02 highlighted the possible effects of muscle regulation during bladder filling and voiding, which are reflected in the larger error volume estimation errors leading to slightly larger LOA within 145ml.

## VI. Discussion

In this workThe performance of estimating bladder volume in vivo was assessed using the wearable BI node ANUVIS and a low complexity BI artefact suppressing algorithm. The operational feasibility of the system was demonstrated in three subjects with healthy bladders using uroflowmetry for the bladder voiding process and urodynamic tests for bladder filling.

Commercially available ultrasound bladder scanners show similar accuracy in bladder volume estimation. The Bladder Scan BVI 3000 showed an absolute error from 29.8 mlto 75.4ml [ 2 4 ] The BVI 9400] showed an error of 132 ± 67 ml in patients with low urine output [25]. In surgical patients with need of catheterization the same device was shown to overestimate the bladder volume by 21.8ml with limits of agreement of -99ml to 140ml [23] when compared with catheter measured urine volume. In the same study, another ultrasound bladder scanner, *Prima*, showed an underestimationof bladder volume by -20.7 ml and wide limits of agreement from -183ml to 141ml.

An average error -29.65 ± 87.6 ml. was observed in the estimate of the total fluid in a full bladder and the residual fluid after voiding for all considered cases in three subjects. Bland Altman analysis for intermittent measurements emulating real time operation showed that the presented sensor node and algorithm had a bias of -5.2 ml with LOA between 119.7 ml to -130.1 ml. This result demonstrated that the volume estimates provided by the algorithm were influenced by an unaccounted external factor. The observed over/under-estimation of bladder volume by the presented algorithm could be due to other confounding factors such as altered muscle tenor [38], leading to a change in the BI of the muscles involved in urine retention and evacuation from the bladder. The observed increase in BI during coughing, as seen in Figure 7 for BF1 and BF2 may be due to the change in the detrusor or pelvic muscle strength or muscle tone while maintaining bladder compliance. A possible effect of muscle clenching was also observed on S02 during bladder filling as seen in (Section III, Figure 7 and Figure 8 of supplementary material.), where an increase in the BI was observed after reaching a volume approximately 260ml. Also, during voiding, an unexplained increase in BI was observed mostly either at the beginning or towards the end of the voiding cycle. Previous research evaluating the feasibility of EIT systems for bladder volume measurement have reported a rise in BI during bladder filling and voiding, with confounding effects of muscle activity[27]. Research in the field of sports medicine has shown that the muscle BI of athletes is larger in magnitude than the general population due to greater muscle strength and muscle tone [39]

The presented algorithm could be further improved with sensor fusion from additional data such as inertial sensors .and electromyography (EMG) sensors and to be able to interpret the change in the measured BI accurately. In this study the aim was to provide a proof of concept demonstrated on three subjects for a non-invasive, autonomous bladder volume measurement system. Additional information from different sensor sources, on the muscle activity could provide necessary information for the onset of changes of major muscle groups involved in urine storage and voiding to provide useful insights into the processes of the pelvicregion [40]. Such data would be crucial for identifying different confounding factors such as increasing BI during muscle contraction/relaxation, the effects of the outflow rate during voiding, and the applied force by the abdomen to evacuate the urine from the bladder. In turn, this information would provide means to develop a model for BI change *only* due to volume changes in the bladder and decouple the artefacts and/or confounding effects due to the involved muscle mechanisms for storing and voiding urine from the bladder.

## VII. Conclusion

An algorithm for bladder volume estimation from BI measurements was developed and executed to provide a proof of concept for the operation feasibility of a non-invasive, automated and context-based system that does not require the presence of skilled personnel. The changes in the BI of the pelvic floor were studied during the processes of bladder filling and voiding using a tetrapolar BI sensor node. The expansion of the urinary bladder was studied with ultrasound measurements and this analysis led to finalizing an electrode placement on the lower abdomen. With the used electrode positions, a linear model of changing BI of the lower abdomen with changing bladder volume was deduced with uroflowmetry and urodynamic tests. This was followed by the development of an algorithm for bladder volume estimation. The algorithm processed the measured BI data, extracted features of interest, and identified and corrected artefacts to finally convert the relative changes in measured BI to milliliters

Intermittent measures for each experiment during bladder filling and voiding showed that the volume estimated by the developed algorithm had a bias of -5.2ml with LOA between 119.7ml to -130.1ml. The variations seen in volume estimation are similar to that observed with handheld and clinician operated ultrasound bladder scanners. However, the presented algorithm in combination with a wearable sensor system is advantageous compared to existing technologies as it is fully automated and does not require the presence of a user to perform the measurement. The major source of the error in the volume estimation is the inability of the algorithm to correctly identify confounding changes in due to a change in abdominal muscle tension. Future work will aim to study the effects of muscle activity in the lower abdomen with additional sensing sources. In addition, further studies are needed to analyze the change in the sensitivity of measurement in different subject groups. In the three subjects that were studied during this work, it was observed that the measured ohms per milliliter change scaled with age. If this holds true, it would be beneficial to have a regression analysis on varied population groups that account for age, gender and abdominal adipose content in a future study with larger sample population.

## APPENDIX

Supplementary Material is attached with this work.

# Supplementary Material

## I. TABLES

### Table A
FEATURES RANKED BY THE MRMR ALGORITHM EVALUATING THE BEST FEATURES FOR IDENTIFYING ARTEFACTS IN BI DATA

| Rank | BI Feature | SE Feature |
|------|------------|------------|
| 1 | Drift flag | SE1 Drift flag |
| 2 | Slope of linear fit | SE2 Energy |
| 3 | Intercept of linear fit | SE3 Gradient of mean |
| 4 | Gradient of mean | SE4 Intercept of linear fit |
| 5 | Standard deviation | SE4 Drift flag |
| 6 | Energy | SE2 Drift flag |
| 7 | Maximum value | SE1 Minimum value |
| 8 | Entropy | SE4 Minimum value |
| 9 | Minimum value | SE2 Entropy |
| 10 | Peak-to-peak value | SE2 Minimum value |
| 11 | Mean | SE1 Maximum value |

### Table B
FEATURES SELECTED BY THE RFE ALGORITHM EVALUATING THE BEST FEATURES FOR IDENTIFYING ARTEFACTS IN BI DATA

| BI Feature name | SE Feature name |
|-----------------|-----------------|
| Standard deviation | SE1 Mean |
| Entropy | SE1 Energy |
| Gradient of mean | SE2 Mean |
| Minimum value | SE3 Mean |
| Maximum value | SE3 Energy |
| Peak-to-peak value | SE4 Mean |
| Slope of linear fit | |
| Drift flag | |

### Table C
RESULTS OF SVM CLASSIFIER

| Property name | Value |
|---------------|-------|
| Overall accuracy | 0.673 |
| Accuracy class 0: No artifact | 0.965 |
| Accuracy class 1: Random fluctuations | 0.916 |
| Accuracy class 2: Positive drift | 0.824 |
| Accuracy class 3: Negative drift | 0.828 |
| Precision | 0.805 |
| Recall | 0.892 |
| F-Score | 0.846 |

### Table D
RESULTS OF DNN CLASSIFIER

| Property name | Value |
|---------------|-------|
| Overall accuracy | 0.842 |
| Accuracy class 0: No artifact | 0.976 |
| Accuracy class 1: Random fluctuations | 0.901 |
| Accuracy class 2: Positive drift | 0.832 |
| Accuracy class 3: Negative drift | 0.826 |
| Precision | 0.841 |
| Recall | 0.834 |
| F-Score | 0.838 |

## II. BLADDER VOIDING CASES

### a. Bladder Voiding Case 3 (BV3): S02

Figure1 shows the change in the BI during the first measurement of BI of the lower abdomen during voiding from S01. The uroflowmeter shows that urine was voided for a duration of 47.6s in total but majority of the urine was voided within the first 40s. The total void volume was 717.4ml with a maximum flow rate of 37.2ml/s achived in the first 5s of voiding. The measured BI increases from 516Ω to 544.4 Ω as urine is voided from the bladder, showing an absolute change of 28.4Ω. On an average the sensitivity to bladder volume change parameter ($\delta$) for this voiding experiment is approximately 0.039Ω/ml. A very rapid initial increase in the BI is observed in the first 5s with an absolute change of 18Ω followed by a relatively gradual increase of 10Ω for the remainder of the voiding.

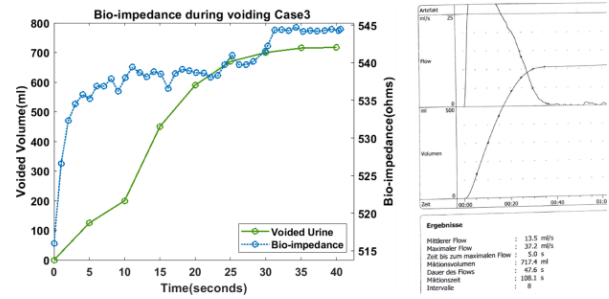

Fig. 1. S02 bladder voiding case3

### b. Bladder Voiding Case 4 (BV4): S02

Figure2 shows the second voiding event recorded with S01. The uroflowmeter reports a total duration of voiding as 27.8s. The total voided volume is 553.6ml and a majority of the urine is voiding within the first 20s. A high maximum flow-rate of 50.7ml/s is recorded that is achieved very quickly within 0.3s of voiding. It is observed that the BI increases from 1113.64Ω to 1135.8Ω with an absolute change of 21.16Ω for the whole voiding process. A sharp increase in BI is observed at t = 1.2s to 2.4s with an absolute change of approximately 7.5Ω. Thereafter, the BI increases continuously till t = 11s where a slight decrease in the BI is observed. This decrease coincides with a rapid decease in the flow-rate of the voided urine volume as measured by the uroflowmeter. On average a sensitivity of 0.038Ω/ml is observed.

### c. Bladder Voiding Case 5 (BV5): S02

Figure3 shows the change in the BI of the lower abdomen during voiding during the third recorded voiding event with S01. The figure on the right is taken from the uroflowmeter and shows that urine was voided for 16.7s in total but the major portion of the urine was expelled from the bladder within the first 10s. A total volume of 317ml was voided with a maximum flow rate of 36.5ml/s that was obtained within the first 5s of voiding. The BI measured increases from 170.7Ω to 184.1 Ω to as urine is voided from the bladder, showing an absolute change of 13.4Ω. On an average the sensitivity to bladder volume change parameter ($\delta$) for this voiding experiment is approximately 0.042Ω/ml. The BI change during measurement follows a constant rate of change in general with slight deviations between 10s to 15s.

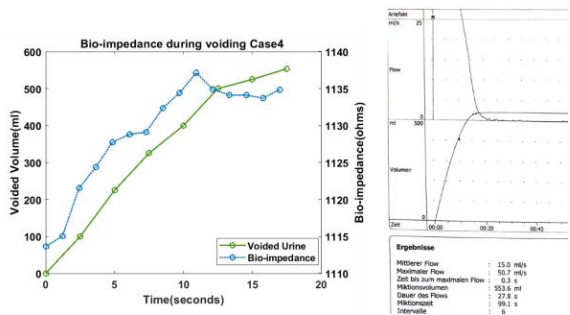

Fig. 2. S02 bladder voiding case4

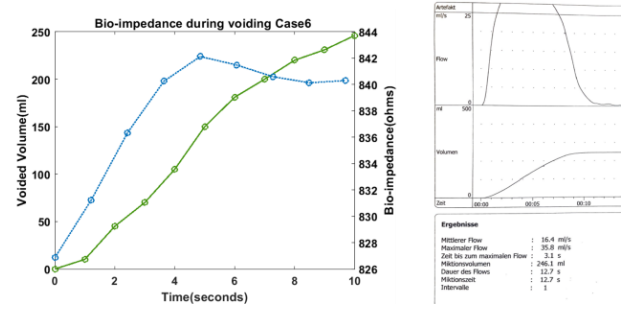

Fig. 4. S02 bladder voiding case6

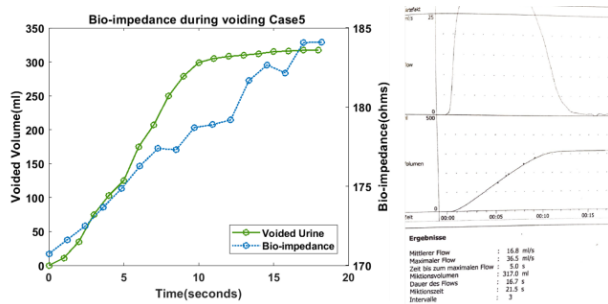

Fig. 3. S02 bladder voiding Case5

### d. Bladder Voiding Case 6 (BV6): S02

Figure4 shows the change in the BI of the lower ab- domen during the fourth voiding event with S01. The uroflowmwter figure on the right shows that the total time for voiding urine was 12.7s in total and a majority was voided in the first 10s. A total volume of 246.1ml was voided with a maximum flow rate of 35.8ml/s that was obtained within the first 3s of voiding. The BI measured increases from 826.8Ω to 840 Ω to as urine is voided from the bladder, showing an absolute change of 13.2Ω. On an average the sensitivity to bladder volume change parameter ($\delta$) for this voiding experiment is approximately 0.054Ω/ml. The BI increases rapidly in the first 5s, after which both the rate of change of BI and the absolute values of BI show a decrease converging to a value of 840 Ω at the end of the voiding cycle.

### e. Bladder Voiding Case 7 (BV7): S03

Figure5 shows the change in the BI of the lower abdomen during voiding. The figure on the right is taken from the uroflowmeter and shows that urine was voided for 32.8s in total but the major portion of the urine was expelled from the bladder within the first 24s. A total volume of 568.4ml was voided with a maximum flow rate of 37.2ml/s that was obtained within the first 5s of voiding. The BI measured increases from 325.8Ω to 350 Ω to as urine is voided from the bladder, showing an absolute change of 25.3Ω. On an average the sensitivity to bladder volume change, shown by the parameter ($\delta$), for this voiding experiment is approximately 0.044Ω/ml. The BI change, however, is not strictly monotonic and it can be seen to rise quickly in the first 2 seconds. Thereafter, the BI increases continuously with an almost constant rate of change.

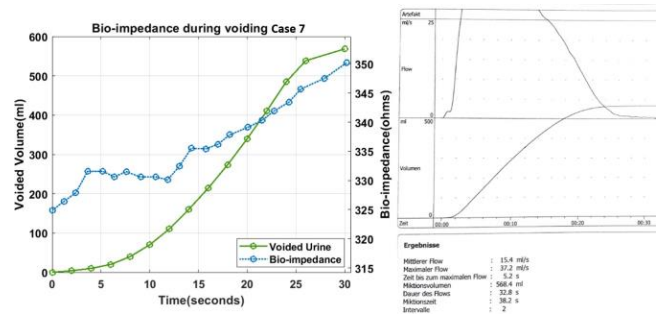

Fig. 5. S03 bladder voiding Case7

### f. Bladder Voiding Case 8 (BV8): S03

Figure6 shows the change in the BI during the second measurement of BI of the lower abdomen during voiding from S03. The figure on the right is taken from the uroflowmeter and shows that urine was voided for 13.7s in total but the major portion of the urine was expelled from the bladder within the first 10s. A total volume of 307.9ml was voided with a maximum flow rate of 41.1ml/s that was obtained within the first 3s of voiding. The BI measured increases from 1298.4Ω to 1312 Ω to as urine is voided from the bladder, showing an absolute change of 13.6Ω. On an average the sensitivity to bladder volume change parameter ($\delta$) for this voiding experiment is approximately 0.044Ω/ml. Here an initial rapid increase in the BI is seen which followed by abrupt lowering of the measured BI around 6s. After 6s, the BI increases although the rate of change varies till the end of the voding.

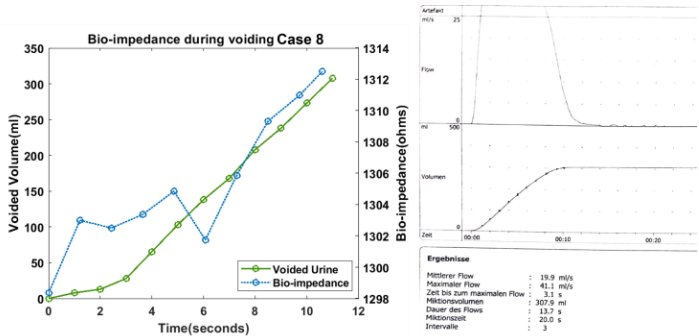

Fig. 6.  S03 bladder voiding Case8

## III. Bladder Filling

### a. Bladder Filling case 3 (BF3): S02 Urodynamic Test1 Fill1
The first cystometry test was done with slower fluid infusion rates starting from 25ml/min to 45ml/min over a period of 14 minutes. A total infused volume of 509ml was recorded. In the first filling cycle as shown in Figure 7, a SDV was marked at t=12 minutes after start of inflow of fluid. As fluid is infused into the bladder, the BI decreases and a change of -13.864Ω from t =0 minutes to t = 8 minutes was seen for an infusion of 285ml. Around t=8 minutes after the start, a growing desire to void was reported by the subject and at this point it was seen that the BI increased with addition of fluid in the bladder.The observed increase in BI of 2.864Ω lasted for the remainder of the fluid infusion. The change in the direction of the BI change started approximately 4 minutes before a strong desire to void(SDV) was expressed.

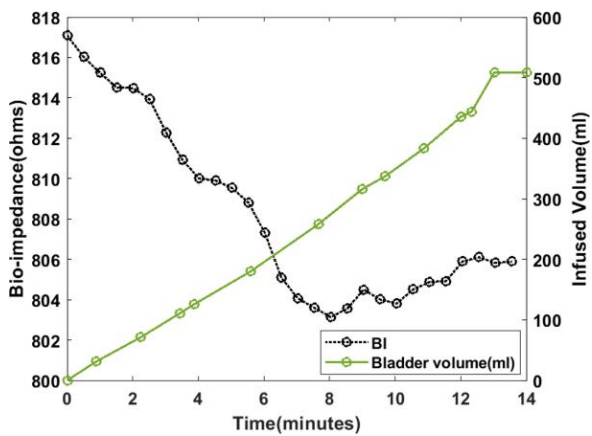

Fig. 7.  S02 bladder filling case3

### b. Bladder Filling case 4 (BF4): S02 Urodynamic Test1 Fill2
The consecutive filling of the bladder was performed with a faster bladder filling rate of 50ml/min on the same day as Case3 with S02. The total bladder volume at the end of the filling cycle was 438ml over a filling period of 9minutes. Similar to the first filling cycle, different attributes such as coughing, SDV, subject talking etc were recorded. SDV was marked at 9 minute of the filling cycle, right at the very end of the filling cycle while FDV was marked at the 5 minute mark.

During the faster filling of the bladder, the BI measured showed more fluctuating data than what was observed in the first filling cycle. Figure8 shows that the BI shows fluctuations from t=0.5 minutes to t=2.5minutes after which the BI decreases monotonically until t=5.5 minutes with an overall decrease of 6Ω. At this point in the test, the bladder volume is approximately 300 ml and the BI starts increasing with continued infusion of fluid into the bladder. Time between the change in BI trend and the SDV is 3.5 minutes in this case.

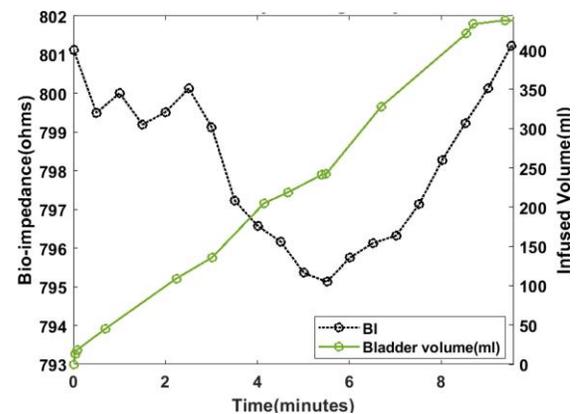

Fig. 8.  S02 bladder filling case 4

### c. Bladder Filling case 5 (BF5): S02 Urodynamic Test2 Fill1
For the second day of testing with S02, the same pattern of infusion rate was followed as previously shown in Case3. In the first filling cycle, fluid was slowly filled into the bladder for a total duration of 13 min and a total volume of 511ml. The SDV was expressed by the subject at t=12 minutes. Till t=8 minutes of the filling cycle, the BI decreased monotonically with increase in fluid as shown in Figure 9. At this point 297ml of fluid was stored in the bladder and the BI changed by -15Ω. The subject reported a tensing of the muscles in the pelvic region at this point. After the t=8 minute mark, the BI started increases, similarly to that seen in Case 3 before. However, during this experiment an increase of 11.454Ω in BI was observed post a bladder volume of approximately 300ml, which is a greater change than that observed in Case 3.

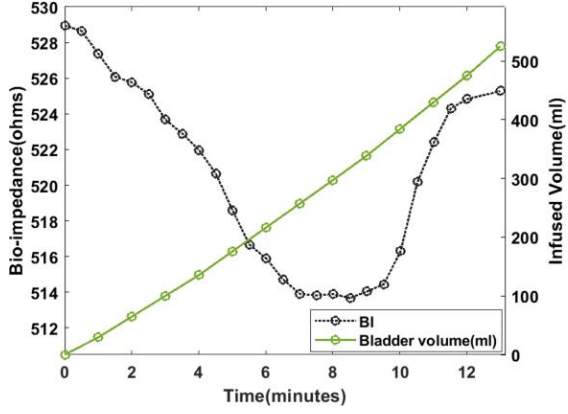

Fig. 9. S02 bladder filling case5

*d. Bladder Filling case 6 (BF6): S02 Urodynamic Test2 Fill2*
The second filling cycle on the day 2 test with S02, was done at a faster rate of 50ml/min infused fluid in the bladder similar to the fill2 cycle on day 1 reported in Case4 .The total bladder volume at the end of the filling cycle was 517.5 ml over a filling period of 10 minutes. The hypothesised decrease in BI was observed only in the first 2 minutes where the BI decreased rapidly by 10Ω. Thereafter the BI increased with increasing bladder volume until it fluctuated again around t=7 minutes. A feeling of pressure or muscle contraction was not reported by the subject to explain these trends, especially in the first two minutes.

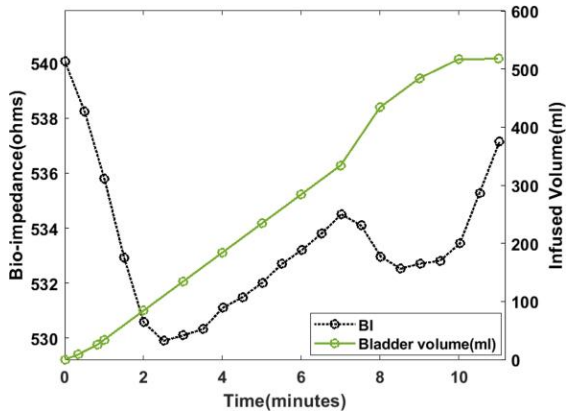

Fig. 10. S01 bladder filling Case6

## IV. BLADDER VOLUME ESTIMATION

In this section, the results of the offline implementation of the bladder volume estimation on each dataset are presented. Datasets from both bladder filling and voiding are presented for subjects S02 and S03.

### A. Bladder Filling (BF)

#### A.1 BF Case3: S02

1) Case 3 of bladder filling showed that as the volume of the bladder increased, the BI decreased continuously till t = 8 minutes where the bladder volume was 285ml and the total measured change in BI was 13.864Ω. After a bladder volume of 285ml, the BI starting increasing till the end of the urodynamic test as shown in Figure11(a).

2) Figure11(b) shows the results of the implemented model for artefact identification and classification. It shows the BI mean of the data windows, the label for positive drift ( Label 2) and the label for negative drift( Label 3). When neither positive nor negative drift occurs, the measured BI indicates no change (Lable 0). As explained before, in the context of bladder filling, Label 0 and Label 2 are considered artefacts. The algorithm correctly detects and classifies the different labels, accurately predicting the onset of positive BI changes after t =8 minutes.

3) The BI estimates ($\hat{BI}$) calculated after suppression of artefacts can be seen in Figure11(c). It can be observed that Kalman filtering is effectively implemented in the data window where artefact (Label 0 and Label 2) is identified.After t=8 minutes, where the BI increases, the filtering method suppresses the observed anomalies. The total $\hat{BI}$ change after artefact suppression is approximately is 16.864Ω.

4) The Kalman Gain (Figure11(d)) value decreases for timestamps where artefact is detected thereby reducing the weight of the measured BI and allowing the predictive model value to dominate over the filter output estimates.

5) Figure11(e) shows the volume estimated against the ground truth provided by the urodynamic test. For the first 8 minutes of bladder filling (upto 285ml), the error in estimated bladder volume is less than 50ml as seen in Figure11(f). However, after the onset of artefacts, the error in volume estimation goes upto 141.1 ml at the end of the test, even after the implemented artefact suppression.

### A.2 BF Case4: S02

1) Case 4 of the urodynamic test for bladder filling had a faster fluid inflow rate (50ml/min) and showed some initial fluctuations in measured BI followed by a decrease in BI. As the volume of the bladder increased the BI decreased by 6Ω till t = 5.5 minutes where the bladder volume was 258ml. After a bladder volume of 258ml, the BI starting increasing till the end of the urodynamic test as shown in Figure12(a).

2) Figure12(b) shows the results of the implemented model for artefact identification and classification with the the mean of the measured BI at each data window, the label for positive drift ( Label 2) and the label for negative drift( Label 3).The algorithm correctly detects and classifies the different labels, accurately predicting the onset of positive

BI changes at t = 0.5minute, t = 1.8minutes and after t =5.5 minutes.

3) $\hat{BI}$ calculated after suppression of artefacts can be seen in Figure12(c). It can be observed that Kalman filtering is effectively implemented when artefact (Label 2) is identified. At the timestamps where artefacts are detected, the estimate $\hat{BI}$ has a lower value as compared to the measured BI value. At t = 5.5minutes, the filter successfully corrects the increasing BI measurements according to the linear rate of change model. Overall, the absolute change in impedance with respect to the first $\hat{BI}$ for the entire duration of the test is 10.23Ω.

4) Figure12(e) shows the volume estimated by the algorithm against the ground truth provided by the urodynamic test. The initial fluctuation in $\hat{BI}$ seen during the first few minutes, impair the results delivered by the algorithm, due to which a consistent under-estimation of the bladder volume is calculated. Consequently, a constantly increasing error in estimated bladder volume is observed with an error of 215.6ml at the end of the bladder filling, as shown in Figure12(f).

*A.3  BF Case5: S02*

1) Case 5 of the urodynamic test for bladder filling followed a slower flow rate for fluid infusion as Case3. The bladder was filled for a total duration of 13 minutes with a total volume of 511ml. Similar to Case3, the rate of decrease in BI measures reduced around t= 7.5 minutes and began increasing after t = 9 minutes at a bladder volume was 297ml as shown in Figure13(a).

2) Figure13(b) shows the results of the implemented algorithm for artefact identification and classification. The algorithm correctly detected and classified the different labels, accurately identifying no change in BI at t=8minutes and the onset of positive BI changes at t = 9minute.

3) The $\hat{BI}$ calculated after the suppression of artefacts can be seen in Figure13(c). Kalman filtering is effectively implemented when artefact, in terms of no change in BI or increasing BI, is identified.

   At t = 7.5minutes, the $\hat{BI}$ corrects the non-changing (and later increasing) measured values of BI according to the linear rate of change model. Overall, the absolute change in impedance with respect to the first $\hat{BI}$ for the entire duration of the test is 24Ω.

4) Figure13(e) shows the volume estimated against the ground truth provided by the urodynamic test. It can be seen that the estimated bladder volume closely matches the true bladder volume initially, followed by an over-estimation starting in the middle section of the experiment as shown in mates the volume by 10.8ml.

*A.4  BF Case6: S02*

1) Case 6 of urodynamic test had the bladder filled at a faster rate (50ml/min). Case 6 showed a rapid decrease in measured BI till t =2.1 minutes with a bladder volume of 120ml and a measured impedance change of 10Ω. Thereafter, the BI increased till t = 7 minutes (with bladder vol- ume 300ml). A decrease in BI measures for a brief period of time was seen at this juncture, after which the BI increased again as shown in Figure14(a).

2) Figure14(b) shows the results of the implemented algorithm for artefact identification and classification. The algorithm correctly detects and classifies the different labels, accurately predicting the onset of positive BI changes at t = 2.1 to 7minutes and at t≥9 minutes.

3) $\hat{BI}$ calculated after suppression of artefacts can be seen in Figure12(c). It can be observed that Kalman filtering is effectively implemented when artefact (label 2) is identified between t = 2.1 to 7 minutes. After t = 7 minutes, a *no artefact* decision is made by the algorithm. $\hat{BI}$ are calculated as described in the artefact suppression section of the main manuscript by adding the ΔBI between consecutive windows to the previous BI estimate.

4) Figure12(e) shows the volume estimated against the ground truth provided by the urodynamic test. The rapid change in the initial BI cause an early onset of overestimation in the volume estimates. The level of the volume estimate changes in accordance to the change in impedance values of $\hat{BI}$, finally resulting in an underestimation of the bladder volume as shown in Figure12(f).

**B.  Bladder Voiding (BV)**

*B.1  BV Case3: S02*

1) The change in measured BI with bladder voiding in S02 is measured in Case3. The total volume  of 717.4ml was voided from the bladder over  40s. An absolute increase of 28.4Ω during this time as seen in Figure15(a).

2) Artefacts are accurately detected in the time series bladder voiding data shown in Figure15(b)with the algorithm where the measured BI does not increase.

3) The Kalman filter corrects the measured BI to provide $\hat{BI}$ (Figure15(c),(d)) according to the Figure13(f). The final bladder volume estimated by the algorithm after Kalman filtering overesti-used model equation at timestamps where arte- facts are detected.

4) (Figure20(e)) shows that the algorithm majorly over-estimated the bladder volume at the begin- ning of the voiding. But the total voided volume estimate provided by the algorithm differs by only 11.346ml from the ground truth as seen in (Figure15(f)).

5) Also in this case, the rapid increase in measured BI at the beginning of the experiment could be due to the contraction of bladder and abdomi- nal muscles responsible for changing the muscle properties and BI, which probably confounded the bio-impedance measures.Similar to that seen in Case1 of bladder voiding, the high voiding flow rate (37.2ml/s) achieved within the first 5s in the uroflowmeter data (Figure1) corroborate this reasoning.

### B.2  BV Case4: S02

1) A total volume of 553.6ml was voided from the bladder with a majority of volume voided in 20s during voiding. An absolute increase of $21.16\Omega$ was observed during this time as seen in Figure16(a).

2) Artefacts are accurately detected with the algo- rithm at timestamps where the measured BI does not increase in the time series bladder voiding data shown in Figure16(b).

3) Artefact correction using Kalman filter sup- presses the measured BI with lower gain values to provide $\hat{BI}$ (Figure16(c),(d))according to the used model equation at timestamps where arte- facts are detected.

4) After volume conversion, (Figure16(e)) shows an underestimation of 13.11ml on the ground truth of the total voided volume. In spite of the low magnitude error on the total voided volume, an over-estimation of the volume is seen from the beginning that results in an error of up to 138ml as seen in (Figure16(f)).

5) A rapid increase in measured BI (impedance change of $7.5\Omega$) at the beginning of the exper- iment between 1.2s to 2.4s could be responsible for the observed over-estimation. As observed in the cases before, this could be due to the con- traction of bladder and abdominal muscles with very high voiding flow rate (50.7ml/s) achieved within the 0.3s according to the uroflowmeter data (Figure2).

### B.3  BV Case5: S02

1) A total volume of 317ml was voided from the bladder with a majority of volume voided in 10s during voiding in Case5. An absolute increase of $13.4\Omega$ was observed during this time as seen in Figure17(a).

2) This voiding instance did not have many artefact instances and the few that were observed were de- tected by the algorithm as shown in Figure17(b) .

3) The estimates $\hat{BI}$ by the Kalman filter corrected the measured BI when artefact was detected as shown in (Figure17(c) and (d))according to the used model equation at timestamps where artefacts are detected.

4) An underestimation of the bladder volume

(Figure17(e)) is seen in this case, especially be- tween 5s to 15s where the error linearly increases upto 119ml(Figure17(f)). However, the final vol- ume estimate is only 26.12ml less than the true bladder volume provided by the uroflowmeter.

5) In this case the uroflowmeter data (Figure3) shows a voiding flow rate of 36.5ml/s achieved within 5s, which is only slightly less than that seen in Case1 (Figure5) and Case3(Figure1). Case5 bladder voiding data does not show rapid increase in BI at the beginning and the BI mea- surements change uniformly through the voiding cycle. Yet, it does not seem to follow the pat- tern of voided volume change provided by the uroflowmeter, resulting in an under-estimation of voided volume for a major portion of the test. Therefore, again, an additional sensing source would be required to understand the different aspects (changing bladder volume, muscle tone, muscle activity etc) affecting the measured BI.

### B.4  BV Case6: S02

1) Case6 of bladder voiding presents the least amount of voided urine volume of 246.1ml in 12.7s. An absolute increase of $13.2\Omega$ was ob- served during this time as seen in Figure18(a).

2) Artefacts, in the form of decreasing BI measures are detected towards the end of the bladder void- ing data shown in Figure18(b).

3) The Kalman filter corrects the measured BI with lower gain values to provide $\hat{BI}$ (Figure18(c),(d)) according to the used model equation at times- tamps where artefacts are detected.

4) The estimated volume (Figure18(e)) shows devi- ations from the ground truth in this case with an overestimation with an error of 85.3ml in total voided volume and a maximum error of upto 200ml (Figure18(f)).

5) Also in this case,the rapid increase in measured BI (approximately $11\Omega$) at the beginning of the experiment (with voiding flow rate (35.8ml/s) achieved within the first 3.1s in the uroflowmeter data (Figure4) could be due to the contraction of bladder and abdominal muscles responsible for changing the muscle properties and BI.

### B.5  BV Case7: S03

1) A total volume of 568.4ml was voided from the bladder i, with a majority of volume voided in the first 24s. An absolute increase of $25.3\Omega$ was observed during this time as seen in Figure19(a).

2) As opposed to the previous cases for bladder fill- ing, Label 3 and Label 0 are considered artefacts for voiding events and are accurately detected by the algorithm in the time series bladder voiding data shown in Figure19(b).

3) At timestamps where artefacts are detected, the Kalman filter corrects the measured

BI with lower gain values to provide $\hat{BI}$

(Figure19(c),(d))according to the used model equation.

4) Finally, (Figure19(e)) shows that the ground truth and the volume estimated by the used algorithm.The error on the total volume of voided urine between the ground truth and algorithm estimated volume differs by 22.128ml. However, errors in estimated volume are observed at the start of the bladder voiding with a maximum error of 150 ml measured at the beginning (Figure19(f)).

5) In this case, high flow rates(37.2ml/s) are achieved within the first 5s as shown in the uroflowmeter data (Figure5). The rapid increase in measured BI at the beginning of the experiment could be due to the contraction of bladder and abdominal muscles (thereby changing the muscle properties and BI) which might confound the bio-impedance measures and hence lead to errors in volume estimation.

### B.6 *BV Case8: S03*

1) In Case8 of bladder voiding, a total volume of 307.9ml was voided from the bladder and most of the volume was voided in the first 10s. An absolute increase of $13.6\Omega$ was observed during this time as seen in Figure20(a).

2) Figure20(b) shows the artefact detection by the algorithm, identifying anomalous behaviour where measured BI starts decreasing.

3) At timestamps where artefacts are detected, the Kalman filter corrects the measured BI with adjusted gain values to provide $\hat{BI}$ (Figure20(c),(d))according to the used model equation.

4) Finally, the volume estimation is implemented (Figure20(e)) which shows that the ground truth and the estimated volume by the used algorithm. The estimate differed by only 1.426ml from the ground truth on the total volume of voided urine. However, higher errors in estimated volume are observed at the start of the bladder voiding with the max error of 90 ml measured within the first 3s as shown in (Figure20(f)).

5) Again as seen previously in Case 1, the errors in volume estimation in this case can be linked to the high urine outflow rate (41.1ml/s) achieved within the first 3s in the uroflowmeter data as seen in (Figure6). The rapid increase in measured BI at the beginning of the experiment could be attributed to the contraction of bladder and abdominal muscles that might lead to a change in the muscle properties and hence the BI, confounding the bio-impedance measures.

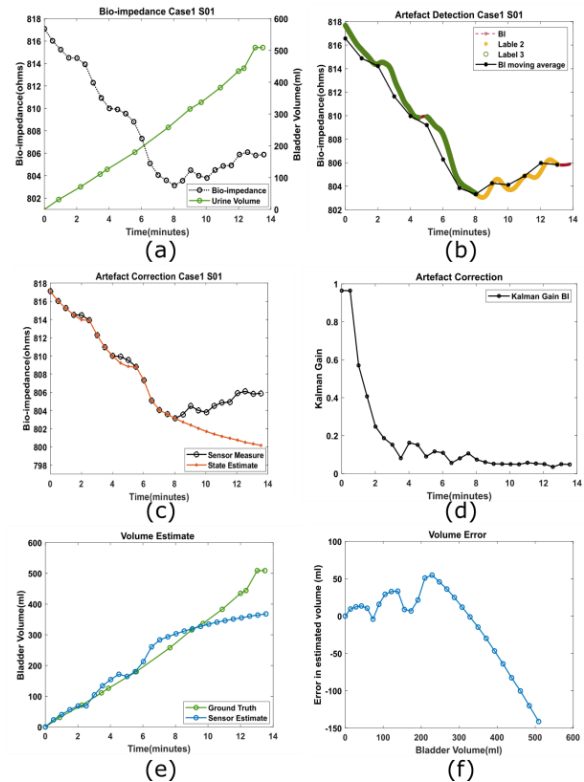

(a) (b)
(c) (d)
(e) (f)

Fig. 11. Bladder volume estimation with artefact identification and correction for Case3 bladder filling(7).(a) Bio-impedance(BI) changes during bladder filling. (b) Identification and classification of data features into labels indicative of presence of artefact in BI data. (c) Result of artefact suppression and temporal data of BI estimates after implementation of Kalman filtering. (d) Temporal values of Kalman gain for each data window. (e) Volume estimate of implemented algorithm against the ground truth provided by urodynamic testing. (f) Error in volume estimation at each data window.

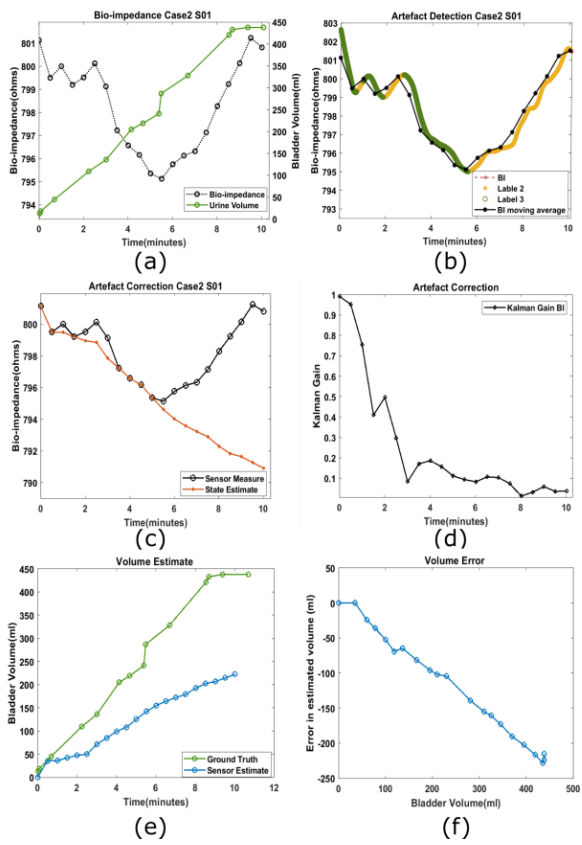

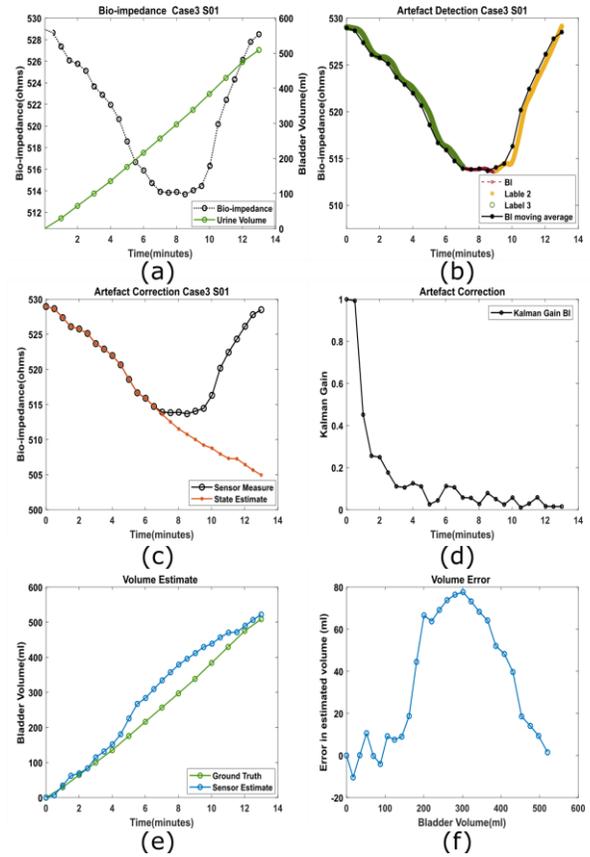

Fig. 12. Bladder volume estimation with artefact identification and correction for Case4 bladder filling(8).(a) Bio-impedance(BI) changes during bladder filling. (b) Identification and classification of data features into labels indicative of presence of artefact in BI data. (c) Result of artefact suppression and temporal data of BI estimates after implementation of Kalman filtering. (d) Temporal values of Kalman gain for each data window. (e) Volume estimate of implemented algorithm against the ground truth provided by urodynamic testing. (f) Error in volume estimation at each data window.

Fig. 13. Bladder volume estimation with artefact identification and correction for Case5 bladder filling(9).(a) Bio-impedance(BI) changes during bladder filling. (b) Identification and classification of data features into labels indicative of presence of artefact in BI data. (c) Result of artefact suppression and temporal data of BI estimates after implementation of Kalman filtering. (d) Temporal values of Kalman gain for each data window. (e) Volume estimate of implemented algorithm against the ground truth provided by urodynamic testing. (f) Error in volume estimation at each data window.

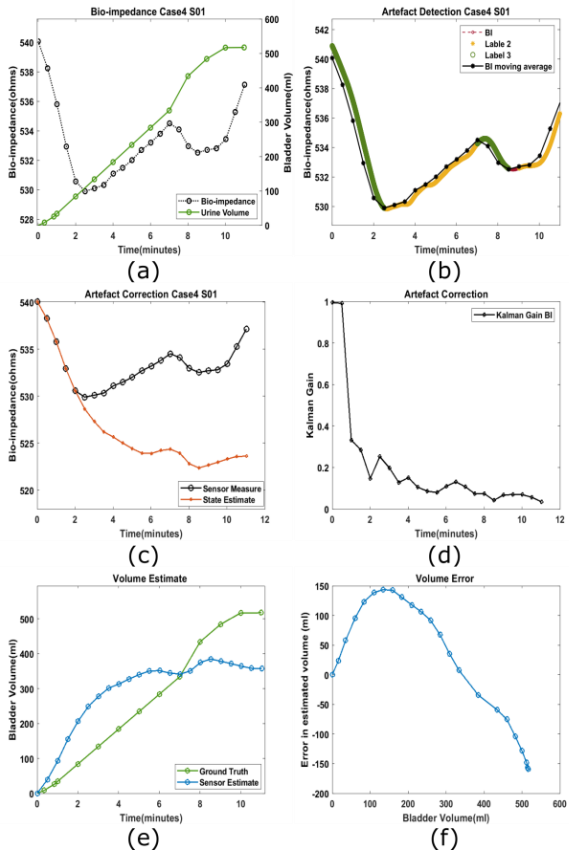

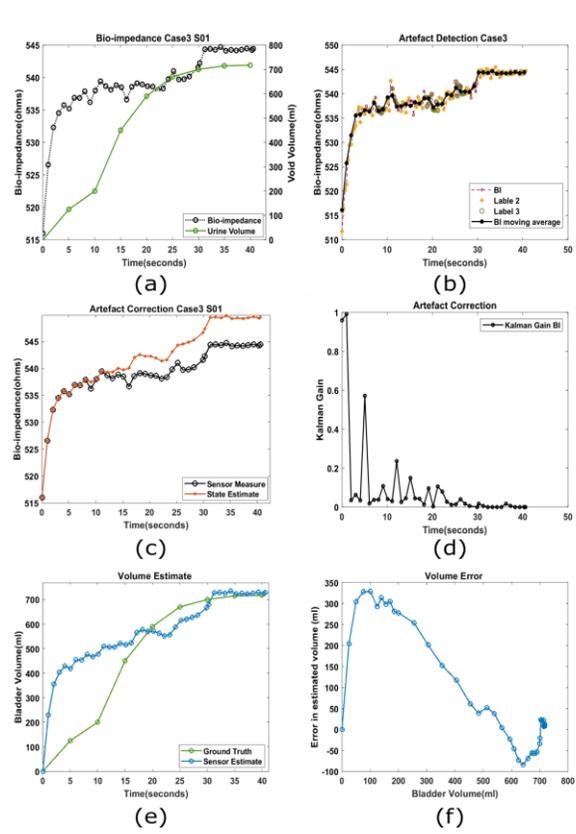

Fig. 14. Bladder volume estimation with artefact identification and correction for Case6 bladder filling(10). (a) Bio-impedance(BI) changes during bladder filling. (b) Identification and classification of data features into labels indicative of presence of artefact in BI data. (c) Result of artefact suppression and temporal data of BI estimates after implementation of Kalman filtering. (d) Temporal values of Kalman gain for each data window. (e) Volume estimate of implemented algorithm against the ground truth provided by urodynamic testing. (f) Error in volume estimation at each data window.

Fig. 15. Bladder volume estimation with artefact identification and correction for Case3 bladder voiding(1).(a) Bio-impedance(BI) changes during bladder voiding. (b) Identification and classification of data features into labels indicative of presence of artefact in BI data. (c) Result of artefact suppression and temporal data of BI estimates after implementation of Kalman filtering. (d) Temporal values of Kalman gain for each data window. (e) Volume estimate of implemented algorithm against the ground truth provided by urodynamic testing. (f) Error in volume estimation at each data window.

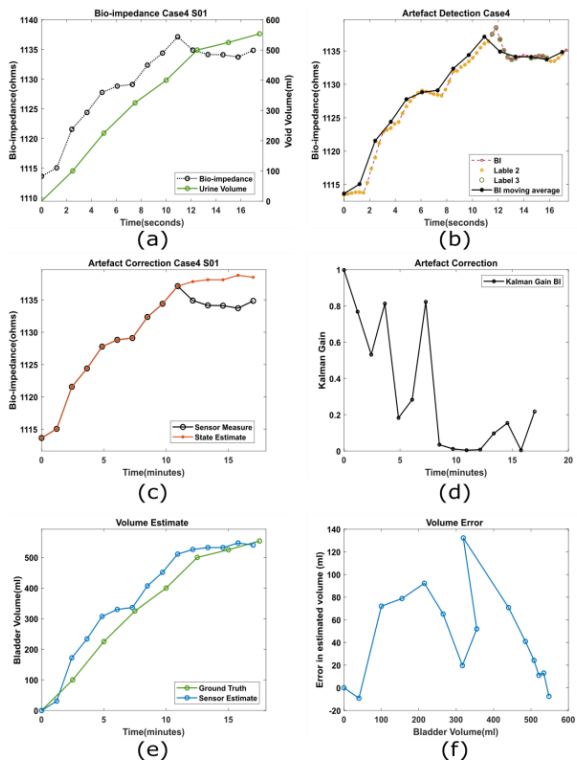

Fig. 16. Bladder volume estimation with artefact identification and correction for Case4 bladder voiding(2).(a) Bio-impedance(BI) changes during bladder voiding. (b) Identification and classification of data features into labels indicative of presence of artefact in BI data. (c) Result of artefact suppression and temporal data of BI estimates after implementation of Kalman filtering. (d) Temporal values of Kalman gain for each data window. (e) Volume estimate of implemented algorithm against the ground truth provided by urodynamic testing. (f) Error in volume estimation at each data window.

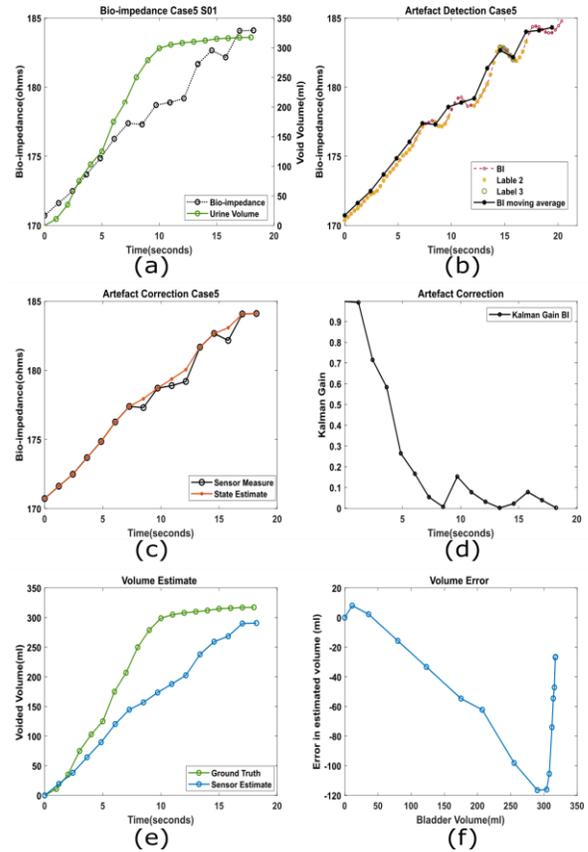

Fig. 17. Bladder volume estimation with artefact identification and correction for Case5 bladder voiding(3). (a) Bio-impedance(BI) changes during bladder voiding. (b) Identification and classification of data features into labels indicative of presence of artefact in BI data. (c) Result of artefact suppression and temporal data of BI estimates after implementation of Kalman filtering. (d) Temporal values of Kalman gain for each data window. (e) Volume estimate of implemented algorithm against the ground truth provided by urodynamic testing. (f) Error in volume estimation at each data window.

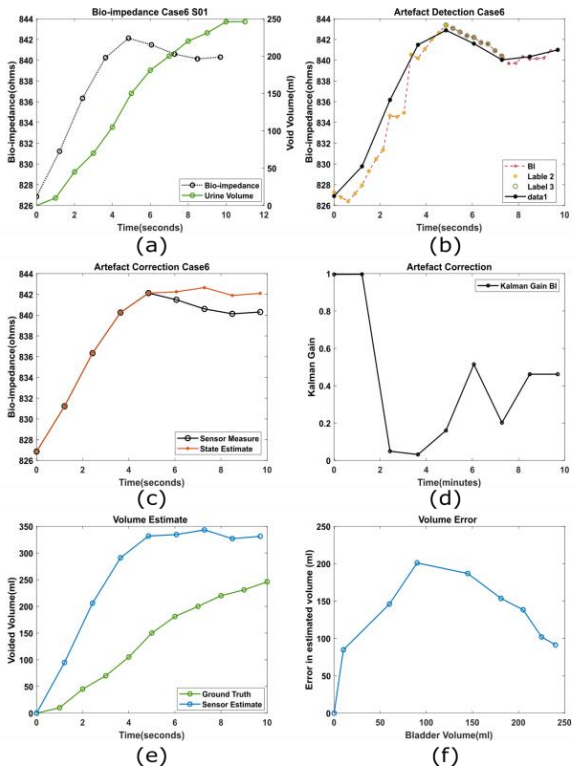

Fig. 18. Bladder volume estimation with artefact identification and correction for Case6 bladder voiding(4).(a) Bio-impedance(BI) changes during bladder voiding. (b) Identification and classification of data features into labels indicative of presence of artefact in BI data. (c) Result of artefact suppression and temporal data of BI estimates after implementation of Kalman filtering. (d) Temporal values of Kalman gain for each data window. (e) Volume estimate of implemented algorithm against the ground truth provided by urodynamic testing. (f) Error in volume estimation at each data window.

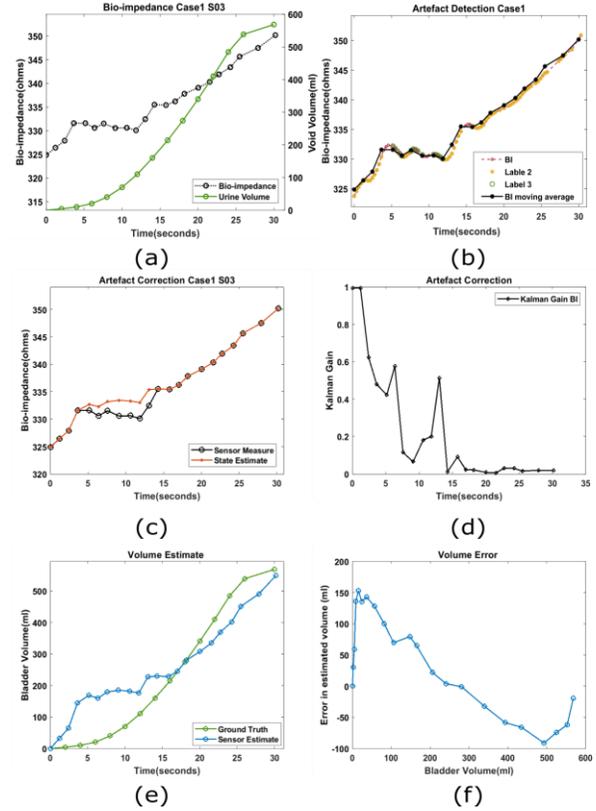

Fig. 19. Bladder volume estimation with artefact identification and correction for Case7 bladder voiding(5).(a) Bio-impedance(BI) changes during bladder voiding. (b) Identification and classification of data features into labels indicative of presence of artefact in BI data. (c) Result of artefact suppression and temporal data of BI estimates after implementation of Kalman filtering. (d) Temporal values of Kalman gain for each data window. (e) Volume estimate of implemented algorithm against the ground truth provided by urodynamic testing. (f) Error in volume estimation at each data window.

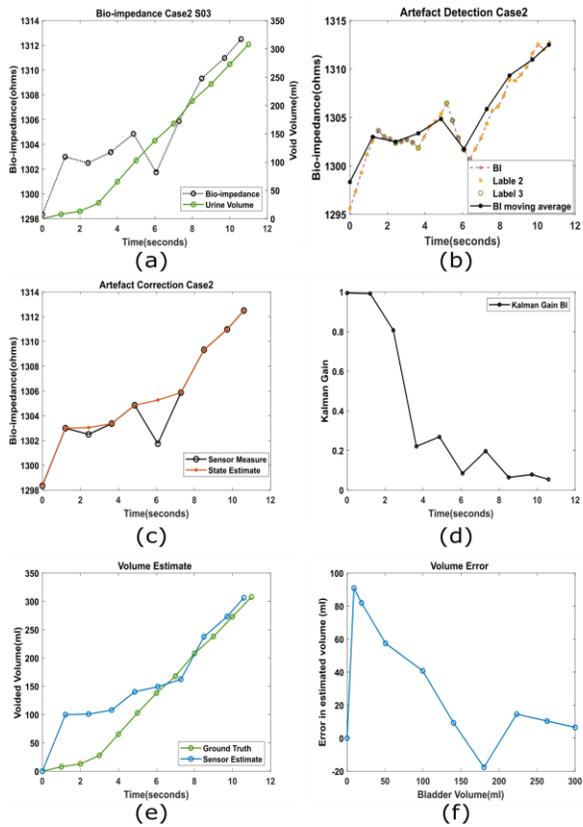

(a)

(b)

(c)

(d)

(e)

(f)

Fig. 20. Bladder volume estimation with artefact identification and correction for Case8 bladder voiding(6).(a) Bio-impedance(BI) changes during bladder voiding. (b) Identification and classification of data features into labels indicative of presence of artefact in BI data. (c) Result of artefact suppression and temporal data of BI estimates after implementation of Kalman filtering. (d) Temporal values of Kalman gain for each data window. (e) Volume estimate of implemented algorithm against the ground truth provided by urodynamic testing. (f) Error in volume estimation at each data window.